\documentclass[12pt]{nature}

\usepackage[utf8]{inputenc}
\usepackage{graphicx,graphics}
\usepackage[outdir=./]{epstopdf}
\usepackage{amssymb}
\usepackage[super,comma]{natbib}
\usepackage{etoolbox}

\newcommand{\irc}{IRC\,+\,10$^{\circ}$216}

\newcommand{\kms}{\mbox{km~s$^{-1}$}}

\newcommand{\msun}{\mbox{$M_{\odot}$}}

\newcommand{\rstar}{\mbox{$R_{*}$}}

\newcommand{\arcsecp}{\mbox{\rlap{.}$''$}} 
\newcommand{\arcsec}{\mbox{$''$}} 
\newcommand{\arcmin}{\mbox{$'$}}

\def\aap{Astron. \& Astrophys.}
\def\aaps{Astron. \& Astrophys. Suppl.}

\def\nat{Nature}

\def\apj {Astrophys. J.}
\def\apjl {Astrophys. J. Lett.}
\def\mnras{Mon. Not. R. Astron. Soc.}
\def\apss{Astrophys. Space Sci.}

\begin{document}
\makeatletter
\renewcommand{\NAT@figcaption}[2][]{
\refstepcounter{figure}
\sffamily\noindent\textbf{Figure \arabic{figure}}\hspace{1em}#2}
\makeatother

\title{Atmospheric molecular blobs shape up circumstellar envelopes of AGB stars} 
\author{L. Velilla-Prieto$^{1}$, J.\,P. Fonfr\'ia$^{2,1}$, M. Ag\'undez$^{1}$, A. Castro-Carrizo$^{3}$, M. Gu\'elin$^{3}$, G. Quintana-Lacaci$^{1}$, I. Cherchneff$^{4}$, C. Joblin$^{5}$, M.C. McCarthy$^{6}$, J.A. Mart\'in-Gago$^{7}$, J. Cernicharo$^{1}$} 
\maketitle
\begin{affiliations}
\item Department of Molecular Astrophysics, Instituto de F\'isica Fundamental (IFF-CSIC), C/ Serrano 121, 28006 Madrid, Spain
\item Centro de Astrobiolog\'ia (CAB), CSIC-INTA, Camino Bajo del Castillo s/n, ESAC campus, 28692, Villanueva de la Ca\~nada, Madrid, Spain
\item Institut de Radioastronomie Millim\'etrique, 300 rue de la Piscine, 38406, Saint Martin d'H\`eres, France
\item Departement Physik, Universit\"at Basel, Klingelbergstrasse 82, 4056, Basel, Switzerland
\item Institut de Recherche en Astrophysique et Plan\'etologie (IRAP), Universit\'e Toulouse 3 - Paul Sabatier, CNRS, CNES, 9 Avenue du Colonel Roche, F-31028 Toulouse, France
\item Harvard-Smithsonian Center for Astrophysics, 60 Garden Street, MA 02138, USA
\item Group of Structure of Nanoscopic Systems, Instituto de Ciencia de Materiales de Madrid (ICMM, CSIC), C/ Sor Juana In\'es de la Cruz 3, 28049 Cantoblanco, Madrid, Spain
\end{affiliations}

\textbf{This is a preprint version of an article published in Nature. 
The final authenticated version is available online at: 
\texttt{https://www.nature.com/articles/s41586-023-05917-9}
DOI: 10.1038/s41586-023-05917-9}

\newpage

\begin{abstract} 
During their thermally pulsing phase (TP), Asymptotic Giant Branch (AGB) stars eject material that forms extended dusty envelopes\citep{2004agbs.book.....H}. 
Visible polarimetric imaging found clumpy dust clouds within two stellar radii of several oxygen-rich stars\citep{2019A&A...628A.132A,2016A&A...591A..70K,2020A&A...635A.200K,2016A&A...589A..91O,2017A&A...597A..20O}. 
Inhomogeneous molecular gas has also been observed in multiple emission lines within several stellar radii of different oxygen rich stars, including W Hya and Mira\citep{2017SciA....3O2149T,2017NatAs...1..848V,2018A&A...620A..75K,2022A&A...660A..94G}. 
At the stellar surface level, infrared images have revealed intricate structures around the carbon semi-regular variable R Scl and in the S-type star $\pi^{\mathrm{1}}$\,Gru\citep{2017A&A...601A...3W,2018Natur.553..310P}. 
Infrared images have revealed as well clumpy dust structures within a few stellar radii of the prototypical carbon AGB star \irc\citep{2006A&A...455..187L,2016MNRAS.455.3102S}, 
while studies of the molecular gas distribution beyond the dust formation zone have also disclosed complex circumstellar structures\citep{2022ApJ...941...90S}. 
Because of the lack of sufficient spatial resolution, however, the distribution of molecular gas in the stellar atmosphere and the dust formation zone of AGB carbon stars is not known, nor is how it is subsequently expelled. 
Here we report observations with a resolution of one stellar radius of the recently formed dust and molecular gas in the atmosphere of \irc. Lines of HCN, SiS, and SiC$_2$ appear at different radii, and in different clumps, which we interpret as large convective cells in the photosphere, as seen in Betelgeuse\citep{2021Natur.594..365M}. 
The convective cells coalesce with the pulsation causing anisotropies that, together with companions\citep{2017A&A...605A.126R,2020Sci...369.1497D}, shape its circumstellar envelope.
\end{abstract}

\begin{figure}
    \centering
    \includegraphics[scale=0.7]{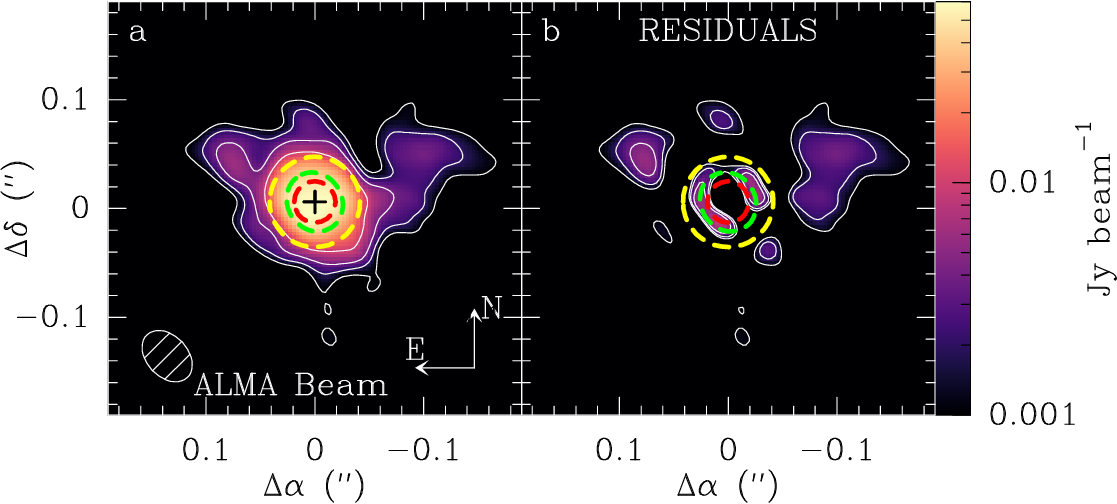}\\
    \caption{Continuum emission of \irc\ at 1\,mm. \textbf{a:} High resolution image of the 1\,mm continuum emission of \irc.
    The contours levels are at 10, 20, 40, and 100\,$\sigma$, where $\sigma$ is the rms of the noise and equal to 92\,$\mu$Jy\,beam$^{-1}$.
    The central cross and the green dashed contour represents the centroid and FWHM of the Gaussian fit to the bulk of the emission.
    The red contour, centred at the position of our estimated Gaussian centroid, shows the photosphere size in the near IR\citep{1988ApJ...326..843R,2001A&A...368..497M}.
    The yellow contour shows the radio photosphere size at 7\,mm from the analysis of high resolution Very Large Array (VLA) observations\citep{2012A&A...543A..73M} .
     \textbf{b:} Map of the residuals of the fit.\label{fig:1}}
\end{figure}

The circumstellar envelope (CSE) of \irc, the archetypal, high-mass-loss-rate carbon AGB star that is also closest to the Sun, was observed with the Atacama Large Millimetre Array (ALMA) interferometer.
These observations spatially resolved the region between 1 and 5\,\rstar\ (that is $\sim$\,20\,--\,100\,mas) with a spatial resolution of 27\,$\times$\,19\,mas.
Further details on the observations and the data processing can be found in the Methods.
The continuum emission map at 1.2\,mm wavelength (Figure\,\ref{fig:1}) shows resolved extensions (above 10\,$\sigma$) that depart from spherical symmetry in addition to a bright central source (see Methods for more details).
A horseshoe shaped structure is seen approximately 25\,mas from the centre of the bright central source (see residuals map in Fig.\,\ref{fig:1}). 
This structure extends from West to South, counterclockwise, and its size is comparable to the stellar size, and could approach that predicted for the convective cells in the upper atmosphere\citep{2017A&A...600A.137F}.
Two other continuum extensions appear further out in Figure\,\ref{fig:1}. 
They are located at approximately 90\,mas\ North-East and 105\,mas\ North-West from the continuum centre and could be dusty gas clumps expelled by \irc, or dust-embedded (sub-)stellar companions \citep{2015A&A...575A..91C,2017MNRAS.471L...1S}. 
We note that such moving clumps have previously been reported in \irc\ and other AGB stars, which we discuss in the Methods.

\begin{figure}
    \centering
    \includegraphics[scale=0.6]{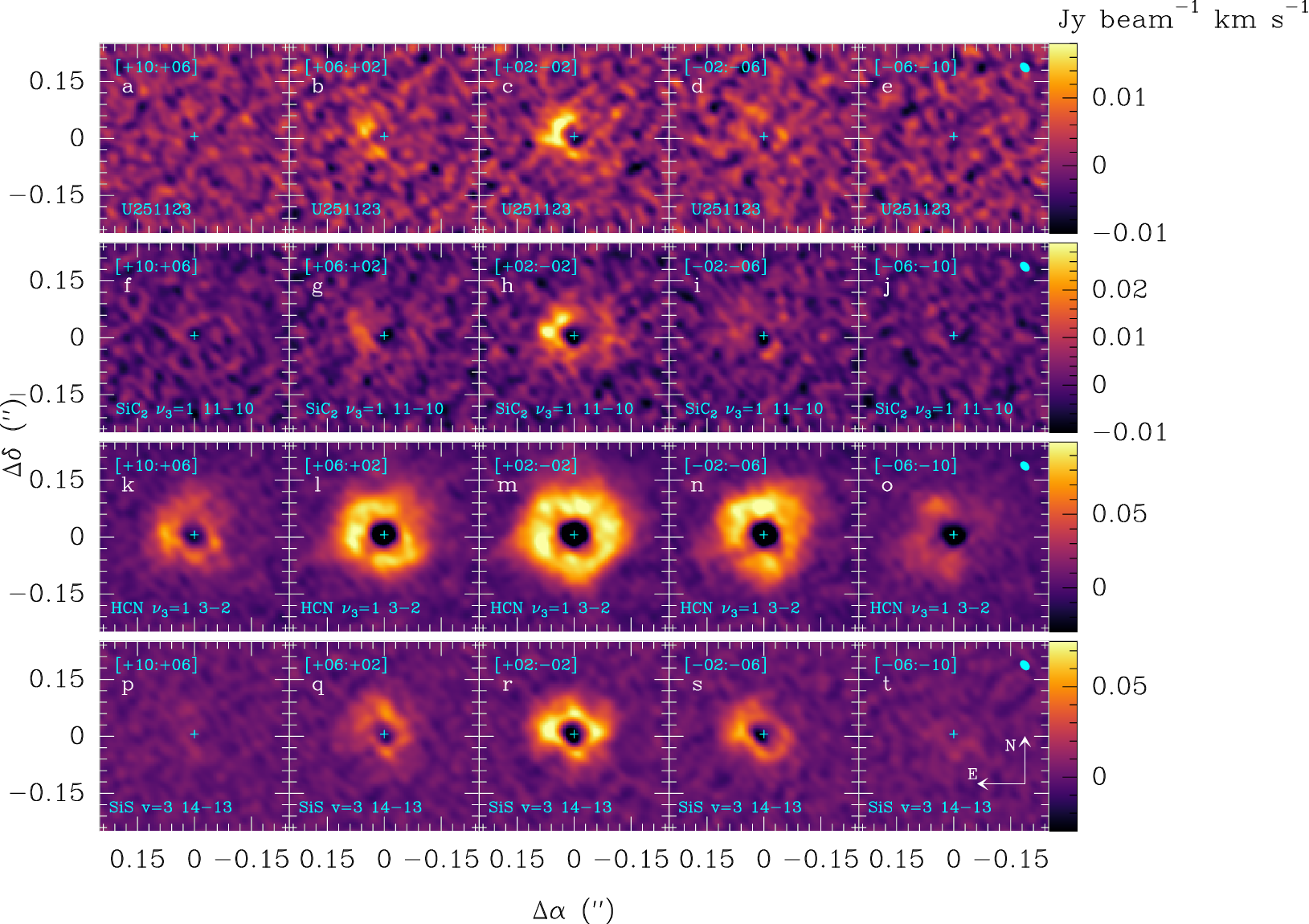}
    \caption{Velocity-integrated maps of different emission lines. \textbf{a-e:} unidentified line U251123 (centroid at 251123\,MHz), \textbf{f-j:} SiC$_2$ $\nu_3$\,=\,1 $J_\mathrm{K,k}$\,=\,11$_{9,3}$--10$_{9,2}$, \textbf{k-o:} HCN $\nu_3$\,=\,1 $J$\,=\,3--2, and \textbf{p-t:} SiS $\nu$\,=\,3 $J$\,=\,14--13. The velocity range where the line has been integrated is shown in the top-left part of each box in \kms\ units. The velocities are relative to the systemic velocity of the source in the LSR, that is -26.5\,\kms\,\citep{2015A&A...575A..91C}. The size and shape of the synthetic beam for each line is shown in the top-right corner of the panels e, j, o, and t. The cross in cyan at the centre represents the position of the peak of the Gaussian fit of the continuum. North is up, East is to the left, as shown in panel \textbf{t}.\label{fig:2}}
\end{figure}

\begin{figure}
    \centering
    \includegraphics[scale=0.85]{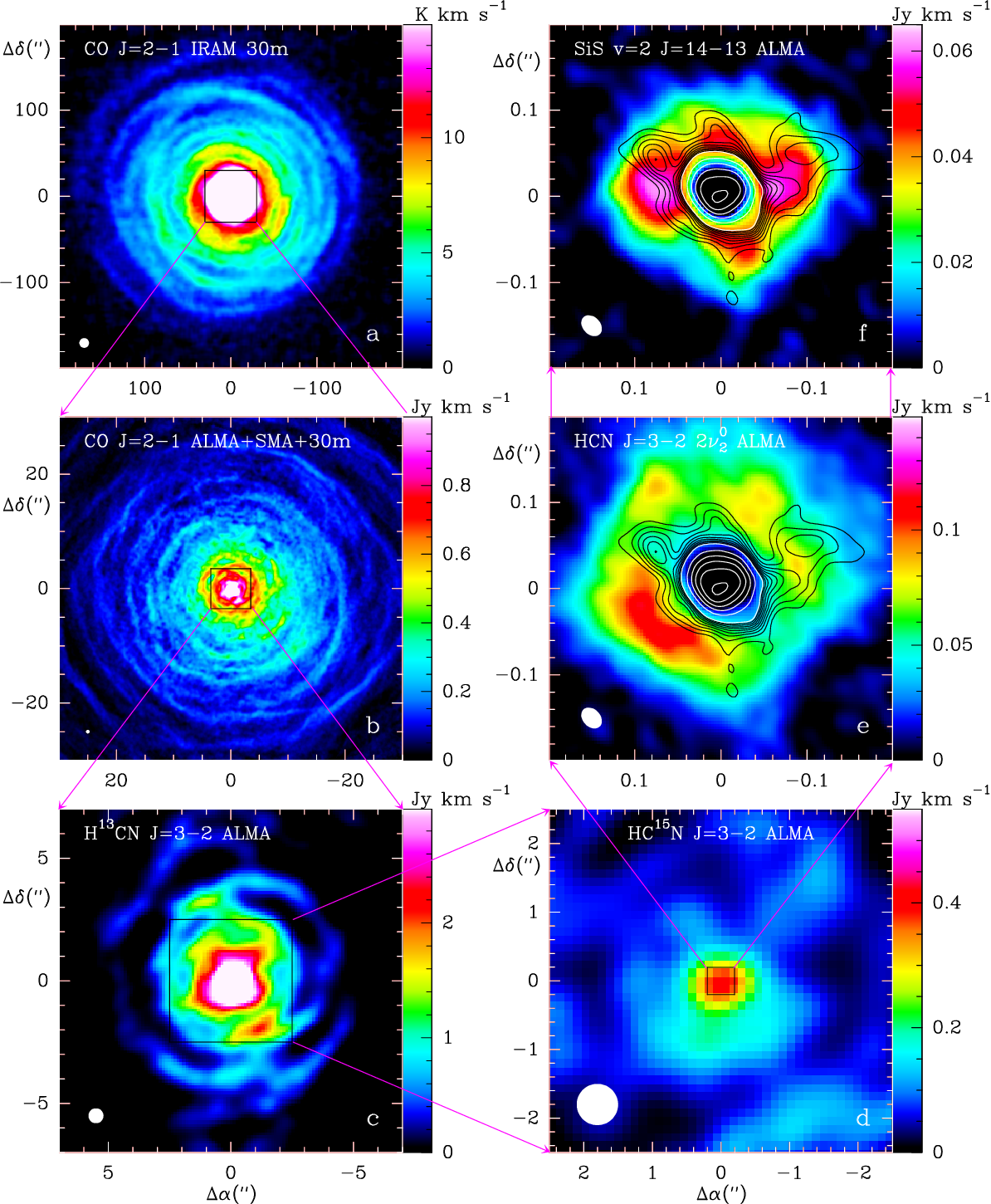}
    \caption{Velocity integrated emission maps of the circumstellar environment of \irc\ for different molecules.
    From the largest scales shown (panels a-b)\citep{2015A&A...575A..91C,2018A&A...610A...4G}, we progressively zoom into the central emission down to scales comparable to a few stellar radii (panels e-f).
    Different telescopes were used to observe the source: the \textit{Institute de Radioastronomie Millim\'{e}trique} (IRAM) 30\,m\ telescope, the Sub-Millimetre Array (SMA), and ALMA.
    The spatial resolution is shown as a white ellipse in the bottom-left corner of each panel.   
    Black and white contours in panels e-f represent, respectively, the continuum emission at levels from 1 to 10 \,mJy in steps of 1\,mJy and from 10 up to 60\,mJy in steps of 10\,mJy.
    1\,mJy is approximately equal to 10\,$\sigma$ (see also Fig.\,\ref{fig:1}). North is up, East is to the left.\label{fig:3}}  
\end{figure}

As concerns the molecular gas, the different molecular species that form show varied and strikingly asymmetric distributions.
Figure\,\ref{fig:2} displays the continuum-subtracted circumstellar emission of SiC$_2$, SiS, and HCN in the atmosphere of \irc\ (see also ED Fig.\,\ref{fig:exdat1} and ED Fig.\,\ref{fig:exdat2}).
The maps show at the position of the star a void of the size of the radio photosphere, that is approximately 40\,mas\ in radius\citep{2012A&A...543A..73M}.
The void is not due to the absence of gas in front of the star but to absorption by foreground gas (see also the Methods).
The line emission arising near the star, which involves upper level energies up to some 6500\,K, clearly deviates from the circular symmetry we would expect in an isotropic formation scenario. 
In particular, emission is almost absent in the Western hemisphere of the envelope for the SiC$_2$ line and the unidentified feature centred at 251123\,MHz (U251123) shown in Figure\,\ref{fig:2}.
Such an asymmetric pattern is commonly observed for other molecular species and takes the form of compact structures and clumps, as well as of emission gaps.
The SiS $v$\,=\,3 $J$\,=\,14--13 emission in the central channels shown in Figure\,\ref{fig:2} displays a very particular pattern of bright spots located North, East, South, and West of the star.
The HCN $\nu_\mathrm{3}$\,=\,1 emission, also shown in Figure\,\ref{fig:2}, displays incomplete arc-like structures, that resemble those seen at much larger scales (see Fig.\,\ref{fig:3}).

Accordingly, both molecular and dust emissions show striking departures from spherical symmetry.
Outside the central region where absorption in front of the stellar continuum occurs, the emission gaps in Figure\,\ref{fig:2} for SiC$_2$ correspond to regions where
the average density, fractional abundances, and temperatures do not meet the required conditions to produce detectable molecular emission.
They hardly may be explained in a scenario of isotropic mass loss.  
This anisotropy in the formation of dust grains and molecules very probably occur as a consequence of localised 
production of matter due to combined pulsation and large-scale convective motions in the photosphere of the star, 
as suggested by Weigelt et al.\citep{1998A&A...333L..51W} and predicted by hydrodynamic models \citep{2015ASPC..497...11C,2017A&A...600A.137F}. 
Episodes of asymmetric ejections of material, leading to the formation of dust clumps in the stellar photosphere of the Red Supergiant star Betelgeuse, have also been proposed to explain its recent Great Dimming \citep{2021Natur.594..365M}.
Montarg\`es et al. (2021)\citep{2021Natur.594..365M} argued that Betelgeuse ejected a bubble of gas some time before the Great Dimming, as a consequence of stellar pulsation in the presence of giant convective cells, which turned into rapid dust formation once the temperature drops below the condensation temperature. 
Similarly, we argue that we are witnessing the anisotropic formation of dust and molecular gas in the high mass-loss rate AGB star \irc\ due to temperature and density anisotropies caused by large convective cells that interact with the stellar pulsation-driven wind\citep{2019A&A...623A.158H}.

\begin{figure}
    \centering
    \includegraphics[scale=0.6]{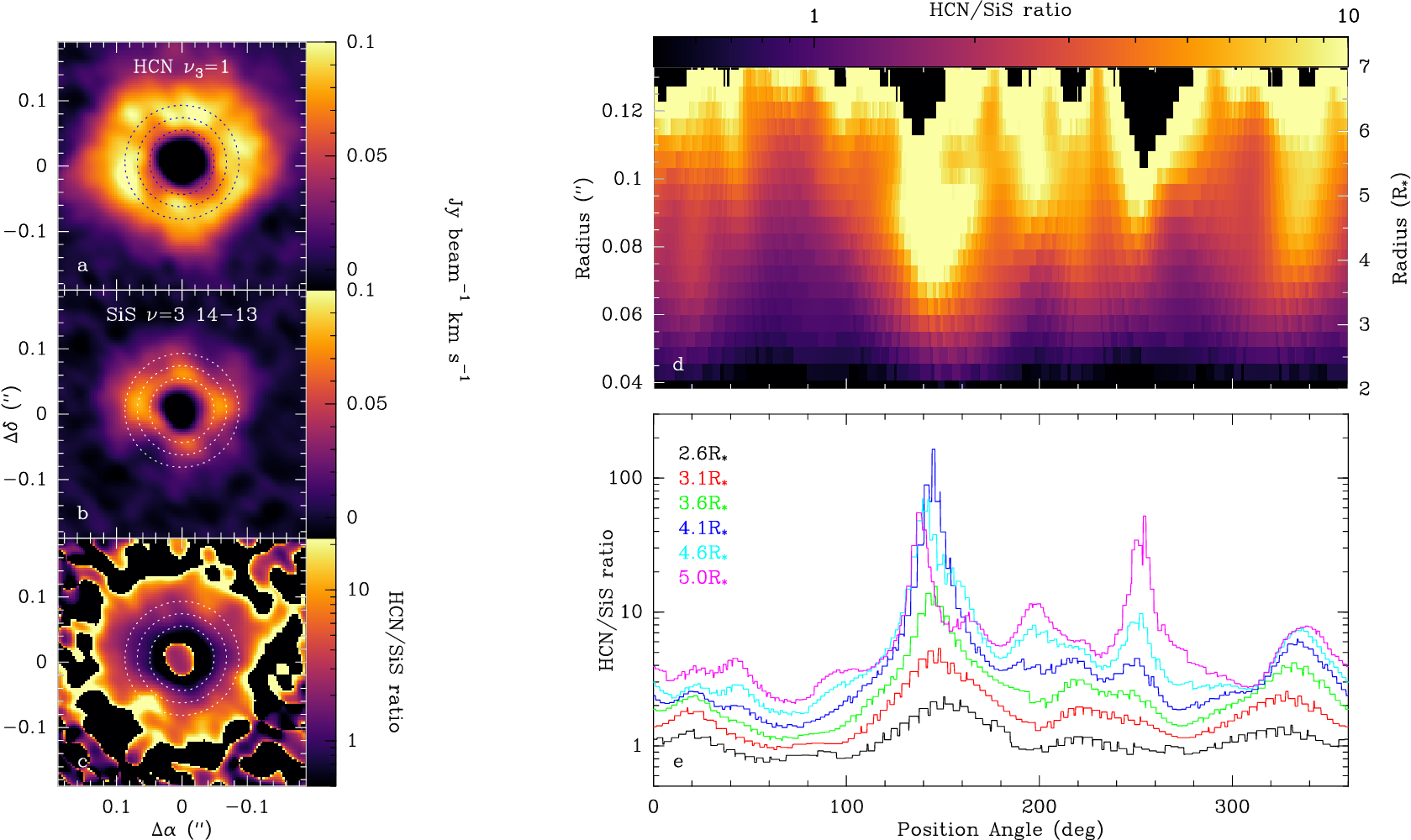}
    \caption{Analysis of the intensity ratio of the HCN $\nu_\mathrm{3}$\,=\,1 $J$\,=\,3--2 and SiS $v$\,=\,3 $J$\,=\,14--13 emission lines. 
    \textbf{a-b:} velocity integrated emission maps of the lines within $\pm$\,2\,\kms\ from the systemic velocity of the source.
    \textbf{c:} map of the intensity ratio between both lines, where the colour scale is in logarithmic scale to improve the visualisation.
    The dotted circles shown in \textbf{a-c} correspond to distances of 2.5, 3.5, and 4.5 stellar radii assuming a 19\,mas radius\citep{1988ApJ...326..843R}.
    \textbf{d:} rectangular projection (position angle versus distance) of the HCN/SiS intensity ratio map to improve the visualisation of the variation of the ratio at different distances. 
    \textbf{e:} curves of the intensity ratio at six different distances from the star.\label{fig:4}}
\end{figure}

In order to get more insight into the causes of the anisotropies observed for HCN and SiS, we have compared the intensity ratio distribution between the SiS $v$\,=\,3 $J$\,=\,14--13 and HCN $\nu_\mathrm{3}$\,=\,1 $J$\,=\,3--2 lines (see Fig.\,\ref{fig:4}).
Both lines have similar upper level energies (3281 and 3042\,K for SiS and HCN lines, respectively). 
Yet, there are large angular variations of their intensity ratio at a given distance from the star.
In particular, the HCN/SiS ratio is strongly enhanced at position angles (PA) 140, 250, and 340$^\circ$, that is South-East, South-West, and North-West.
The fact that, for a given radius, the HCN/SiS line ratio experiences such angular variations implies that the anisotropies in the emission are not solely due line excitation changes caused by fluctuations in the gas density, in which case the fractional abundances of both HCN and SiS would change equally and the ratio would not vary with angle. 
It is likely that they result from anisotropies in the temperature that weight differently on the fractional abundances of HCN and SiS. 
Such behaviour is predicted by chemical equilibrium models, for which we show results in the Extended Data (ED Fig.\,\ref{fig:exdat3} and ED Fig.\,\ref{fig:exdat4}) and describe in the Methods. 
We note that chemical equilibrium models do effectively predict an enhanced sensitivity of the chemical composition to the temperature at the photosphere, where temperature and pressure are very high, while shocks models, remain inherently uncertain, due to both a high sensitivity to the underlying physical structure and an incompleteness of the chemical kinetics network.
This incompleteness relates to the limited knowledge of formation and destruction chemical routes, the restricted database of available reactions rates, and the range of explored temperatures.
Our chemical equilibrium models indicate that the formation of each molecule is favoured over a given pressure-temperature domain. 
For example, a temperature increase reduces the SiS abundance more drastically than for HCN. 
Even a moderate variation of 10\% in the temperature among different regions of the stellar atmosphere can lead to orders of magnitude variations in the fractional abundances of molecules such as HCN and SiS, as well as of dust precursors such as C$_2$ and SiC$_2$ (see ED Fig\,\ref{fig:exdat3}).
Laboratory simulations have shown that the chemistry of C and C$_2$ in the presence of H$_2$ leads to the growth of carbon clusters and the formation of amorphous carbon nanoparticles \citep{2020NatAs...4...97M}. 
On the other hand SiC$_2$ is suspected to be a precursor for the formation of silicon carbide \citep{2015ApJ...806L...3C}.  
Phenomena such as large convection cells at the surface of the star that add up to the pulsation-driven wind should produce variations in the temperature with angle, which in turn cause fractional abundance gradients, and ultimately lead to the anisotropies seen in the molecular emission.

Alternative scenarios considered to explain the asymmetries observed in \irc\ are the effect of a binary companion and shocks in the atmosphere of the star.
\irc\ is very probably part of a multiple star system that shapes up the characteristic shell structure of its outer envelope. 
Its orbital period or periods has or have been estimated by different authors to be $\sim$\,700\,yr\ and 50\,yr\citep{2015A&A...575A..91C,2015A&A...574A...5D,2018A&A...610A...4G}. 
However, the structures we see at atmospheric scales are only 2\,--\,4\,yr\ old, much too young for invoking the gravitational pull by any companion star or planet.
Moreover, manifestations of the binary interaction affecting the close circumstellar environments are not observed in our maps, such as a disk or a torus-like structure or preferred ejection directions. 
Photochemical effects enhanced by the UV emission of a close companion, as recently studied by Siebert et al.\citep{2022ApJ...941...90S}, cannot explain the asymmetries we observe at atmospheric level in the brightness distribution of SiS or HCN either, since several emission voids surround the star.  
We have also found no evidence of shocks, which would have left signatures in the emission maps such as infalling material and narrow velocity features in the emission profiles (see Methods). 
Nevertheless, we cannot rule out the existence of shocks, which, 
in any case, should be triggered by the stellar pulsation and convective motions in the photosphere\citep{2019A&A...623A.158H}.
Further discussion about alternative scenarios and models considering the combination of anisotropic mass loss and binary interaction are presented in the Methods.

The anisotropies reported here change our views on the formation of molecules and dust in the upper atmosphere of the archetypal AGB star \irc, with different molecular species selectively appearing depending on local physical conditions.
Conversely, this has a critical importance for the mechanisms that later drive the winds of intermediate mass stars and for ensuing circumstellar chemistry\citep{2006MNRAS.367.1585I,2012Natur.484..220N,2018A&A...616A.106V}.
We conclude that, independently of the effect of a binary companion in re-shaping the outer circumstellar envelope, the molecular and dust anisotropies observed in \irc\, and presumably present in most late-type stars, lead to the formation of physical structures that, expanding with the stellar wind, shape up the global spatial structure and evolution of the envelope.

\bibliographystyle{sn-standardnature}

\begin{methods}\label{methods}
\subsection{ALMA Millimetre-wavelength interferometric observations, calibration, and reduction}
The observations presented here correspond to the ALMA project 2018.1.01485.
They were carried out during two executions of 67 min each on 2019-07-09 at 17:26:15 (UTC) and 2019-07-10 at 15:44:10 (UTC).
The total integration time on \irc\ was 68 min.
The array was set into a configuration using 43 antennas that achieved 903 different baselines ranging between 111.2\,m and 13.9\,km (configuration c43-9).
This configuration provided an angular resolution of $\sim$20\,mas and a maximum recoverable scale for the most extended emission of $\sim$450\,mas.

\irc\ was observed at sky elevations ranging between 45$^\mathrm{\circ}$ to 54$^\mathrm{\circ}$ on 2019-07-09, and between 28$^\mathrm{\circ}$ to 40$^\mathrm{\circ}$ on 2019-07-10, through single-pointing integrations.
On average, the weather conditions were good, with an atmospheric precipitable water vapour content of approximately 1.2 and 2.1\,mm, humidity between 10 and 25\%, ambient temperatures below 2$^\mathrm{\circ}$\,C, and system temperatures between 80 and 180\,K.
Four spectral windows of 1.875\,GHz bandwidth were centred at the sky frequencies 264.901, 266.701, 249.901 and 251.701\,GHz, with spectral resolutions of 0.55, 0.55, 0.59 and 0.58\,\kms\ (nominal channel width is 488.281\,kHz).

Calibration was performed with the standard pipeline of the Common Astronomy Software Applications (CASA) software package \citep{2007ASPC..376..127M}, and self-calibration and image synthesis with the Grenoble Image and Line Data Analysis Software (GILDAS) package.
The bright calibrator J0854\,+\,2006 was observed to calibrate the bandpass, and also set for absolute flux reference, with fluxes of 2.61, 2.60, 2.69 and 2.68 Jy for each spectral window.  
Observations were carried out in single polarisation mode and only the horizontal (H) polarisation was observed.
The used flux calibration reference is known to be polarised (A.\,Castro-Carrizo, priv. comm.) though the CASA pipeline did not correct the adopted flux by the value corresponding to the H polarisation at the time of the observations.
Due to this, a later correction of  7-10\,\%\ was introduced in the data based on the polarisation knowledge of the flux calibrator, also considering the hour angle during the observations (thanks to the Northern Extended Millimetre Array, NOEMA, observatory database).
Otherwise, the monitoring made by ALMA in the dates around the observations at different frequencies make these flux values very reliable (with an uncertainty $\lesssim$\,5\,\%).
J1002+1216 was observed every 2 minutes to calibrate amplitude and phase gains over time.

The standard phase calibration was subsequently improved by self-calibration on the compact continuum emission of the source.  
Continuum results were obtained per baseband after line subtraction, and then averaged to produce a single continuum map.  
Imaging restoration was made with uniform weighting. 
The used cleaning methods during the image synthesis were Hogbom \citep{1974A&AS...15..417H} and Steer-Dewdney-Ito (SDI)\citep{1984A&A...137..159S}, where the latter is best suited to reduce clumpiness by the identification of clean component groups. 
The SDI method was only used for line emission imaging, and its results were assessed by comparison with the more standard Hogbom ones. 
The obtained mean synthetic beam for the different maps is 27\,$\times$\,19\,mas (with PA equal to 47$^\mathrm{\circ}$; more details can be seen in the caption of each figure).

\subsection{Continuum main component fit and proper motion} 
The position of \irc\ determined as the centre of the Gaussian fit to the bulk of the continuum emission is $\alpha_\mathrm{J2000.0}$\,=\,09$^\mathrm{h}$\,47$^\mathrm{m}$\,57$^\mathrm{s}$.456\,$\pm$\,0$^\mathrm{s}$.002 and $\delta_\mathrm{J2000.0}$\,=\,+13$^\mathrm{\circ}$.16\arcmin43\arcsecp906\,$\pm$\,0\arcsecp033. 
The uncertainty of the position is the result of the addition in quadrature of the Gaussian fit uncertainty and one fifth of the synthesised beam.
The position uncertainty of interferometric observations can be approximated by $\sim$\,0.5\,$\theta_\mathrm{b}$/SNR, that is the beam size divided by ten for a good signal-to-noise ratio (SNR) detection \citep{1988ApJ...330..809R}, that is SNR\,$\geq$\,5. 
For observations with an extended configuration of the interferometer, as it is our case, the accuracy in the position is poorer than that amount by a factor of two due to the atmospheric phase fluctuations. 
Therefore, we adopted a more conservative value for the uncertainty as one fifth of the half power beamwidth (HPBW) synthesised beam.
The integrated flux, within a square box of 400\,mas length centred at the fitted position, amounts to 496\,$\pm$\,11\,mJy, of which some 90\,\%\ corresponds to the bulk of the emission fitted by our Gaussian (see below).
This flux compares well, considering the source variability, with previous measurements at similar wavelengths \citep{1997Ap&SS.251..247L,2014MNRAS.445.3289F,2018A&A...610A...4G}.
The peak intensity of our continuum source is 68.3\,$\pm$\,2.4\,mJy\,beam$^{-1}$ with a rms of the noise (1\,$\sigma$\, level) equal to 92\,$\mu$Jy\,beam$^{-1}$.
The fit to the continuum emission yields a major axis full width at half maximum (FWHM) of 56.0\,$\pm$\,2.7\,mas\ and a minor axis FWHM of 49.9\,$\pm$\,2.3\,mas\ with a position angle of 58\,$\pm$\,26$^\mathrm{\circ}$.

Very accurate measurements of the position of \irc\, carried out with the most extended configuration of the Very Large Array (VLA) were presented and discussed by Menten et al\citep{2012A&A...543A..73M}. 
They reported the following coordinates for \irc: $\alpha_\mathrm{J2000.0}$\,=\,09$^\mathrm{h}$\,47$^\mathrm{m}$\,57$^\mathrm{s}$.4255\,$\pm$\,0$^\mathrm{s}$.0006 and \\$\delta_\mathrm{J2000.0}$\,=\,+13$^\mathrm{\circ}$.16\arcmin43\arcsecp815\,$\pm$\,0\arcsecp010, which represent an offset of $\Delta\alpha$\,=\,458\,$\pm$\,31\,mas and $\Delta\delta$\,=\,91\,$\pm$\,34\,mas with respect to our measured position.
The time span between our observations and those by Menten et al.\cite{2012A&A...543A..73M} is of 13 years.
Menten et al.\citep{2012A&A...543A..73M} analysed available measurements of the position, leading to determine a proper motion of 35\,$\pm$\,1\,mas\,yr$^{\mathrm{-1}}$ and 12\,$\pm$\,1\,mas\,yr$^{\mathrm{-1}}$, in right ascension and declination, respectively.
Considering the time span, the proper motion of the source would match quite well with our right ascension but not with the declination, where our 91\,mas offset is significantly far from the 156\,mas that one would expect even within uncertainties.
A deeper investigation on this issue is out of the scope of this paper, but we suggest that the mismatch might be a consequence of the resulting orbital motion of \irc, if it is part of a binary or multiple system, and the differences in the parallax among observations registered at different epochs.

\subsection{Continuum extended emission} 
We discuss here previous observations of continuum extensions at different infrared (IR) and optical wavelengths connected to the NE and NW extensions seen in our continuum map,
which were also interpreted as dust patches expelled by \irc\ and/or dust embedded (sub-)stellar companions.
The radius of the stellar photosphere, as measured in the infrared (IR), is of 19\,mas, that is $\sim$\,2.5\,AU\ \citep{1988ApJ...326..843R,2001A&A...368..497M}.
We note, for the conversion to physical scales, that the distance to \irc\ has been measured as 123\,$\pm$\,14\,pc \citep{2012A&A...543L...8G}.
It is also worth noting that these continuum IR and optical studies yield only 2-D images, 
whereas molecular line maps yield velocity information which may teach us about the wind acceleration and position along the line-of-sight of the expelled clumps.

Weigelt et al.\citep{1998A&A...333L..51W} reported, from a 2\,$\mu$m\ Speckle interferometry map with the 6\,m Special Astrophysical Observatory (SAO), the presence of at least three dust clumps, the brightest of which is located at 200\,mas NE and 140\, NW from what they assumed the star was.
They argued that the matter ejection from the photosphere is inhomogeneous due to magnetic activity and large convection cells of size 0.8\,R$_\mathrm{*}$ \citep{1997svlt.work..316F}.
This result was followed up by a time monitoring of \irc\ in the near-infrared by Weigelt et al.\citep{2002A&A...392..131W}. 
The authors reported changes in the appearance of the dust shell when comparing different time periods, showing structural variations compatible with the movement and formation of new dust clumps.
Monnier et al.\citep{2000ApJ...543..861M} reported, from observations with the Berkeley Infrared Spatial Interferometer (ISI), an increase in size of the mid-IR source by a factor of two over ten years, 
which they interpreted as a sharp decrease of the mass-loss rate, but it could be linked to the sporadic expulsion of clumps.
Tuthill et al.\cite{2000ApJ...543..284T} reported, from a 2\,$\mu$m\ Keck aperture-masking interferometry observations at four different epochs covering one pulsating cycle, 
the presence to the N and NE of the central source of four dusty knots moving outward at about the 15\,\kms\ molecular shells expansion velocity, 
perhaps indicating an acceleration with increasing radius.
Le{\~a}o et al.\citep{2006A&A...455..187L,2007ASPC..378..309L} reported, from 1.2\,$\mu$m\ and 2\,$\mu$m\ carried out with the Very Large Telescope Nasmyth Adaptative Optics System (VLT/NACO) observations with an angular resolution of 70\,mas, three bright clumps NW, N, and NE of the central
source, reminiscent of those of our continuum image, but located farther.
Kim et al.\citep{2021ApJ...914...35K} reported observations with the Hubble Space Telescope (HST) at three different epochs in 2001, 2011, and 2016 covering the wavelength range between 5800 to 9800\,\AA, 
that show bright knots around the central star.
One of the knots was tentatively identified as a companion star\citep{2015ApJ...804L..10K}. 
Nevertheless, this hypothesis has been dropped in their new study where it is assumed that the knots are dust knots illuminated by starlight through holes in the dusty inner 
envelope layers. 

\subsection{Chemical model}
The chemical model is based on the code presented in Ag\'undez et al\cite{2020A&A...637A..59A}.
We compute the chemical equilibrium composition of gas over a pressure-temperature parameter space relevant for the conditions of an AGB atmosphere (5\,$\times$10$^{-4}\,$--\,5\,mbar and 2000--3000 K). 
Chemical equilibrium is solved by minimisation of the Gibbs free energy of the system. 
We consider 919 gaseous species, including atoms and molecules, with thermochemical data taken from different compilations, and an elemental composition typical of C-rich AGB stars \citep{2020A&A...637A..59A}. 
We show results only for the relevant molecules discussed in this work, that is HCN, SiS, and SiC$_2$, as well as the carbon dust precursor C$_2$ in ED Fig.\,\ref{fig:exdat3} \&\,ED Fig.\,\ref{fig:exdat4}.

\subsection{Shocks, outflows, and binarity} 
The occurrence of shocks in the extended atmospheres of AGB stars is thought to exist as a consequence of the
interplay between the stellar pulsation and the convective motions of material in the outermost shells of the star\citep{2017NatAs...1..848V,2019A&A...623A.158H}.
The periodic ejection of material leads to the creation of a dynamical atmosphere where infalling material, that is still gravitationally bound to the star, could encounter recently ejected material causing temperature and density enhancements across the stellar atmosphere.
We calculated the molecular emission of \irc\ with the aid of the radiation transfer code MADEX \citep{2012EAS....58..251C} for a time dependent envelope model that includes stellar pulsation.
This envelope model is based on the output of the models presented by Cherchneff et al.\citep{2012A&A...545A..12C}, where physical conditions and chemical abundances were computed at different pulsation phases. 
Shocks produced by the pulsation movement of the star, lift outwards the gas layers close to the stellar surface and then the material falls back due to stellar gravity after the shock passage.
The effect of shocks should be seen in the line profiles at the current angular resolution of 20\,mas as narrow features in the profiles occurring at very specific velocities (see ED Fig.\,\ref{fig:exdat5}).
These narrow emission and absorption features appear due to a combination of effects that include the amount of gas and its relative position, velocity, and temperature.
However, as we show in ED Fig.\,\ref{fig:exdat6}\,\&\,ED Fig.\,\ref{fig:exdat7}, the observed line profiles along different strips through the stellar atmosphere lack of any supporting evidence of shocks in any of the species analysed.
Nevertheless, we cannot rule out the existence of shocks that are not seen in our data due to effects such as excitation, gas turbulence, and opacity effects or in the case that they are smoothed in our data due to insufficient resolution.
In ED Fig.\,\ref{fig:exdat8}\,\&\,ED Fig.\,\ref{fig:exdat9} we also show different patterns of emission and absorption depending on the line of sight, where no signatures of shocks are seen.
Inside the stellar disk, we see molecular absorption lines, as a consequence of absorption against the stellar continuum and self-absorption.
Between the photosphere and the radio photosphere, we observe little or no emission and absorption.
Finally, as we look beyond the radio photosphere, only molecular emission is seen.

On the other hand, we have also investigated if there is any emergent bipolar or collimated outflow noticeable in the position-velocity diagrams along different directions.
Such outflow would be most likely caused by a binary companion orbiting the AGB star, as suggested in the formation scenarios of pre-planetary nebulae\citep{1998ApJ...496..833S}.
The presence of a binary companion has been argued to explain different signatures seen in the circumstellar shells of \irc, such as off-centred shells and spiral structure\citep{1993A&A...280L..19G,2015A&A...575A..91C,2015A&A...574A...5D}.  
Kim et al.\citep{2015ApJ...804L..10K} reported the detection of a point-like source in HST images lying 0\arcsecp5 to the SE from \irc\ that could be the companion star. 
Different studies have reported an elongation along the NE-SW direction of \irc's envelope, which would mark a preferential direction or have a role of symmetry axis, and the observation of bright IR clumps moving in this direction \citep{2000ApJ...543..284T,2019A&A...629A.146V} (and references therein).
This has led to suggestions that \irc, which is known to be a star in the latest stages of the AGB phase, is already evolving towards a pre-planetary nebula and starting a bipolar outflow due to the presence of a binary companion. 
Such an outflow would certainly break the symmetry and homogeneity of the stellar wind. 
Nevertheless, we have found no evidence in our data of outflows, disk or torus-like structures, or preferred ejection directions.

Decin et al.\citep{2015A&A...574A...5D} presented sub-arcsecond resolution ($\sim$0.5\arcsec) ALMA data of \irc\ where they suggest the presence of spiral structure in its inner wind. 
We can directly compare that to our maps of HCN in the main text where a more complex structure is seen. 
A clear spiral pattern with complete arms is not seen but some blobs or clumps are clearly noticed. 
Our maps show a much more complex situation where different molecules trace different structures. 
At atmospheric level, it is hard to interpret the spatial distribution of SiS, HCN, and SiC$_2$ seen in our data as a sole consequence of the presence of a companion.

Hydrodynamical models of the interplay between an AGB star and a binary companion\citep{1998ApJ...497..303M,1999ApJ...523..357M} revealed a recognisable degree of symmetry in the patterns generated in the circumstellar material that we do not observe. 
Decin et al.\citep{2015A&A...574A...5D} also presented a 3D model of the envelope with a binary companion where several parameters related to the orbital properties of the system were tested out.
From their analysis, they ruled out the possibility of a Jupiter-like or brown dwarf secondary based on arcsec scale signatures in the position-velocity diagrams. 
However, they assumed a relatively high mass of 4\,\msun\ for \irc, but more recent estimates based on oxygen isotopic ratios would point out to a mass of 1.6\,\msun\citep{2017A&A...600A..71D}. 
Kim et al.\citep{2012ApJ...759...59K} explored the impact that wide binaries have on the morphology of AGB circumstellar circumstellar envelopes.
These models predict the emergence of a spiral pattern that can be seen as a ring-like pattern if the system is observed edge-on (inclination of 90$^\circ$), which seems incompatible with the large scale morphology of \irc's circumstellar envelope\citep{2015A&A...575A..91C,2018A&A...610A...4G}.
Malfait et al.\citep{2021A&A...652A..51M} also presented Smooth Particle Hydrodynamic models of binary interaction with AGBs. 
They explored models of a 1.5\,\msun\ AGB accompanied by a 1\,\msun\ star at 6\,AU (semi-major axis). 
The secondary induces a motion of the stars around the centre of mass of the system while generating a spiral shock, and perturbs the AGB wind material. 
Different patterns can be observed in their models, from perfect spirals to more irregular configurations but none of their simulations resemble our observations. 
Moreover, it would still be very difficult to reconcile the predicted density profile with the chemical segregation that we observe.
Similar models were also presented by Maes et al.\citep{2022arXiv220612278M}. 
Recently, Aydi \& Mohamed\citep{2022MNRAS.513.4405A} have presented 3D hydrodynamical models of the circumstellar environments of evolved stars affected by the presence of nearby sub-stellar companions. 
They concluded that if the companion is at a distance larger than four stellar radii, a single spiral arm should be seen, which can be ruled out from our data. 
For a much closer separation to the companion, a more complex structure of multiple spirals is observed, whose configuration depends on many orbital and system parameters that were explored. 
Nevertheless, we note again that the stellar outflow at the atmospheric level is not an isotropic wind, as it can be seen in the images we present. 
These authors discuss the models of Freytag, Liljegren \& H\"ofner\citep{2017A&A...600A.137F}, which show that convection cells and fundamental-mode pulsation will result in chaotic shock wave structures around the AGB star, shifting from spherical symmetry. 
These structures cause anisotropies in the temperature and density conditions leading to chemical variations in the gas depending on the angle (direction) from the centre of the star. 
We propose that such chaotic structures cause the chemical anisotropies seen in our maps, and that, in combination to the interaction with a binary companion, shape the circumstellar envelope of this archetypal AGB.

In order to test if blobs launched from the photosphere along random directions can explain our observations, we have developed several models that are based on the evolving envelope model presented by Cernicharo et al.\citep{2015A&A...575A..91C}.
Our code is flexible, so it can be used to model an isolated star or a binary system involving an AGB star with different physical and orbital parameters, such as masses, inclination, and period, among others. 
The envelope is formed by a modulated ejection of matter in a spherical wind, which simulates the pulsation-driven outflow in AGB stars, that is the steady-state wind. 
Different contrasts between minima and maxima in the pulsation stage, up to 100, and different periods, up to 3 years, were tested out.
The main development of the model is that the AGB steady-state wind is modified by discrete spherical caps that are randomly ejected.
These spherical caps were modelled as temporary enhancements of the wind density at the photosphere, keeping constant any other physical or chemical quantity for the sake of simplicity.
These caps were centred at random positions across the photosphere with a formation rate probability typically lower than 10--30\%, and a typical angular width of 60$^{\circ}$.
The formation of each cap lasts about 2.5\,months and only one cap is formed at a time.
The time evolution of the system is calculated over a period of time and the radiation transfer problem is solved for the CO $J$\,=\,2--1 line assuming local thermodynamic equilibrium (LTE) to simplify calculations. 
Finally, the produced maps were convolved with a Gaussian beam to simulate the constraints imposed by interferometers to the observations.
We checked with the CASA modelling tool simalma that the array configuration does not significantly modify the region of the envelope explored in the current work.
We note that the code aims to give qualitative information about the formation of structures in the envelope.
More accurate studies are out of the scope of the current work.
The resulting models can be seen in ED Fig.\,\ref{fig:exdat10}.

First, we have developed a ``control'' model, a scenario that should represent the case of an isolated AGB star with a steady-state wind (see case a in ED Fig.\,\ref{fig:exdat10}).
This model generates a smooth circularly symmetric brightness distribution with decaying emission with distance to the star besides the absorption in the innermost region.
In the second scenario we add the presence of a close companion based on the basic parameters presented by Decin et al.\citep{2015A&A...574A...5D}, that is an inclination of 60$^{\circ}$, a separation of 25\,AU and masses of 4\,\msun\  for the AGB and 1.1\,\msun\ for the companion (case b in ED Fig.\,\ref{fig:exdat10}).
In this case, a smooth circularly symmetric outflow is observed in the innermost region of the envelope map plus a spiral pattern, that is generated in the presented image at the North-East and continues revolving clockwise.
We note that we explored a large parameter space including different masses for the AGB and the companion, the separation between the stars, and the inclination of the system.
None of these models produce arcs and blobs as we do observe in the maps of the circumstellar emission. 
These features only appear under a different scenario, the third case (case c in ED Fig.\,\ref{fig:exdat10}), which adds up the randomly ejected blobs to the AGB outflow.
These models generate random blobs at the stellar photosphere that leave anisotropic features in the emission maps with a notable resemblance to those seen in the ALMA maps, 
such as the two blobs seen to the West in ED Fig.\,\ref{fig:exdat10} (c). 
Moreover, arcs are also observed in these models as a consequence of the evolution of the launched blobs and the interaction with the circumstellar material. 
In this case, we have shortened the distance between the stars to 10\,AU\ in order to keep the value of the orbital period as low as possible considering that the kinematic age of the random ejections is of just a few years.
The spiral structure due to the binary companion is not always clearly seen, probably as a consequence of the interaction of blobs with the material trail. 
Nevertheless, we recall that we have not seen any spiral pattern in the observed emission maps.
Modifications of the pulsation contrast between shells and the pulsation period for values below 3\,years, do not change the results. 
We note that the IR light period of \irc\ is about 1.8\,years and in such case any high contrast shell is smoothed out in the outflow by the angular resolution provided by ALMA.
Considering a gas expansion velocity of approximately 5\,\kms\ close to the photosphere, an angular resolution as high as 7\,mas should be necessary for this task.
After all this analysis we can conclude that the anisotropic features only appear if the ejection of blobs is included in the models.

\makeatletter
\apptocmd{\thebibliography}{\global\c@NAT@ctr 37\relax}{}{}
\makeatother

\end{methods}

\begin{addendum}
\item[Data availability]
All the observational products here used are public and available through the ALMA, SMA, and IRAM-30m archives: \\
\texttt{https://almascience.nrao.edu/aq/} \\
\texttt{https://lweb.cfa.harvard.edu/cgi-bin/sma/smaarch.pl} \\
\texttt{https://iram-institute.org/science-portal/data-archive/} 

\item[Code availability]
The 2012 version of the MADEX code, including spectroscopic information for approximately 3500 species, is available as an executable in: \\
\texttt{https://nanocosmos.iff.csic.es/madex/} \\
The chemical equilibrium and the evolving envelope models are available upon request.
Inquiries about the codes can be addressed to the corresponding authors and to the developers of the codes:
Dr.\,M.\,Ag\'undez for the chemical code, Dr.\,J.\,Cernicharo for the MADEX code, and Dr.\,J.\,Cernicharo and Dr.\,J.P.\,Fonfr\'ia for the models of the evolving envelope.

\item [Acknowledgements]
This publication is part of the ``I+D+i'' (research, development, and innovation) projects PID2020-117034RJ-I00, PID2019-107115GB-C21, PID2019-106110GB-I00, and PID2019-105203GB-C22, supported by the Spanish ``Ministerio de Ciencia e Innovaci\'on'' MCIN/AEI/10.13039/501100011033.
This work was also supported by the European Research Council under Synergy Grant ERC-2013-SyG, G.A. 610256 (NANOCOSMOS).
This work was based on observations carried out with the IRAM, SMA and ALMA telescopes. 
 IRAM is supported by INSU/CNRS (France), MPG (Germany) and IGN (Spain). 
 The Sub-millimetre Array is a joint project between the Smithsonian Astrophysical Observatory (USA) and the Academia Sinica Institute of Astronomy and Astrophysics (Taiwan) and is funded by
the Smithsonian Institution and the Academia Sinica. 
This paper makes use of the ALMA data: ADS/JAO.ALMA\#2013.1.01215.S, ADS/JAO.ALMA\#2013.1.00432.S, and ADS/JAO.ALMA\#2018.1.01485.
ALMA is a partnership of ESO (representing its member states), NSF (USA) and NINS (Japan), together with NRC (Canada), NSC and ASIAA (Taiwan) and KASI (Republic of Korea), in cooperation with the Republic of Chile. The Joint ALMA Observatory is operated by ESO, AUI/NRAO and NAOJ.
We have made use of the CASA (Common Astronomy Software Applications) software (\texttt{https://casa.nrao.edu/}).
The GILDAS package was used to reduce and analyse the data (\texttt{https://www.iram.fr/IRAMFR/GILDAS}).
We acknowledge the GILDAS software team, in particular to S.\,Bardeau, for their help and assistance.
We also thanks to IRAM staff to have provided relevant information for the flux calibration from the NOEMA observatory database.
We want to thank the anonymous referees and the editor, Dr. L.J.\,Sage, for the discussion and improvement of this paper.

 \item[Author contributions]
 L.V.-P. led this publication, analysed and interpreted the data, wrote the article, and made most of the figures of the article.
 J.P.F. delivered the method to produce the input data for the radiative transfer models from the shock models output, and developed and produced the models of the evolving envelope.
 M.A. created the chemical code and the associated images.
 A.C.-C. calibrated and reduced the ALMA data.
 M.G. contributed subtantially to the revision of the literature work and the detailed revision of the text.
 I.C. provided the input abundances that were used to produce the shock models.
 J.C. developed the MADEX and the evolving envelope codes, led the ALMA proposal, and contributed to the overall analysis of this work.
 All authors contributed to the conception, design, and writing of the ALMA proposal and to the discussion and revisions of the article.

 \item[Competing Interests] The authors declare that they have no competing financial interests.

 \item[Correspondence] Correspondence and requests for material should be addressed to Luis Velilla-Prieto (email: l.velilla@csic.es, ORCID: 0000-0001-8275-9341) and Jos\'{e} Cernicharo. (email: jose.cernicharo@csic.es, ORCID: 0000-0002-3518-2524).

\end{addendum}

\newpage

\pagebreak

\pagebreak

\pagebreak

\pagebreak

\setcounter{figure}{0}

\textbf{EXTENDED DATA}\\
\begin{figure}
    \centering
    \includegraphics[scale=0.6]{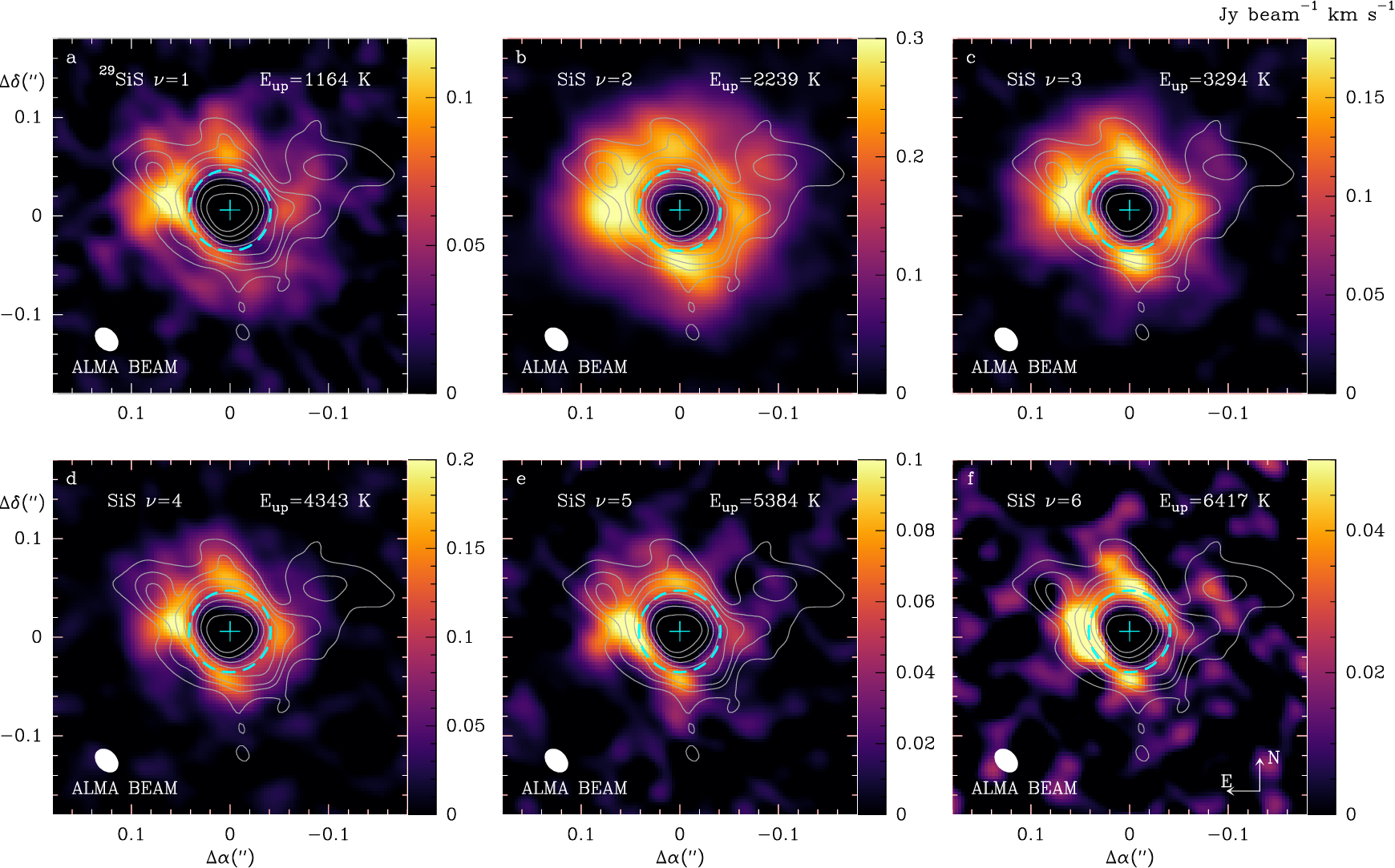}
    \caption{\textbf{Extended data: }Velocity-integrated emission maps of SiS isotopologues in different vibrationally excited states. The velocity-integrated emission maps are shown in colour scale. Note that the intensity scale varies within boxes for improved visualisation. 
    The grey contours correspond to 10, 30, 50, 100, 200, 300, 400, 500, and 600\,$\sigma$ of the rms of the noise of the continuum emission, which is equal to 92\,$\mu$Jy\,beam$^{-1}$.
    The cyan dashed contour represents the radio photosphere \citep{2012A&A...543A..73M}.
    The shape of the synthetic beam is shown in the bottom-left corner of each box.
    North is up, and East to he left.\label{fig:exdat1}}
\end{figure}
\pagebreak

\begin{figure}
    \centering
    \includegraphics[scale=0.6]{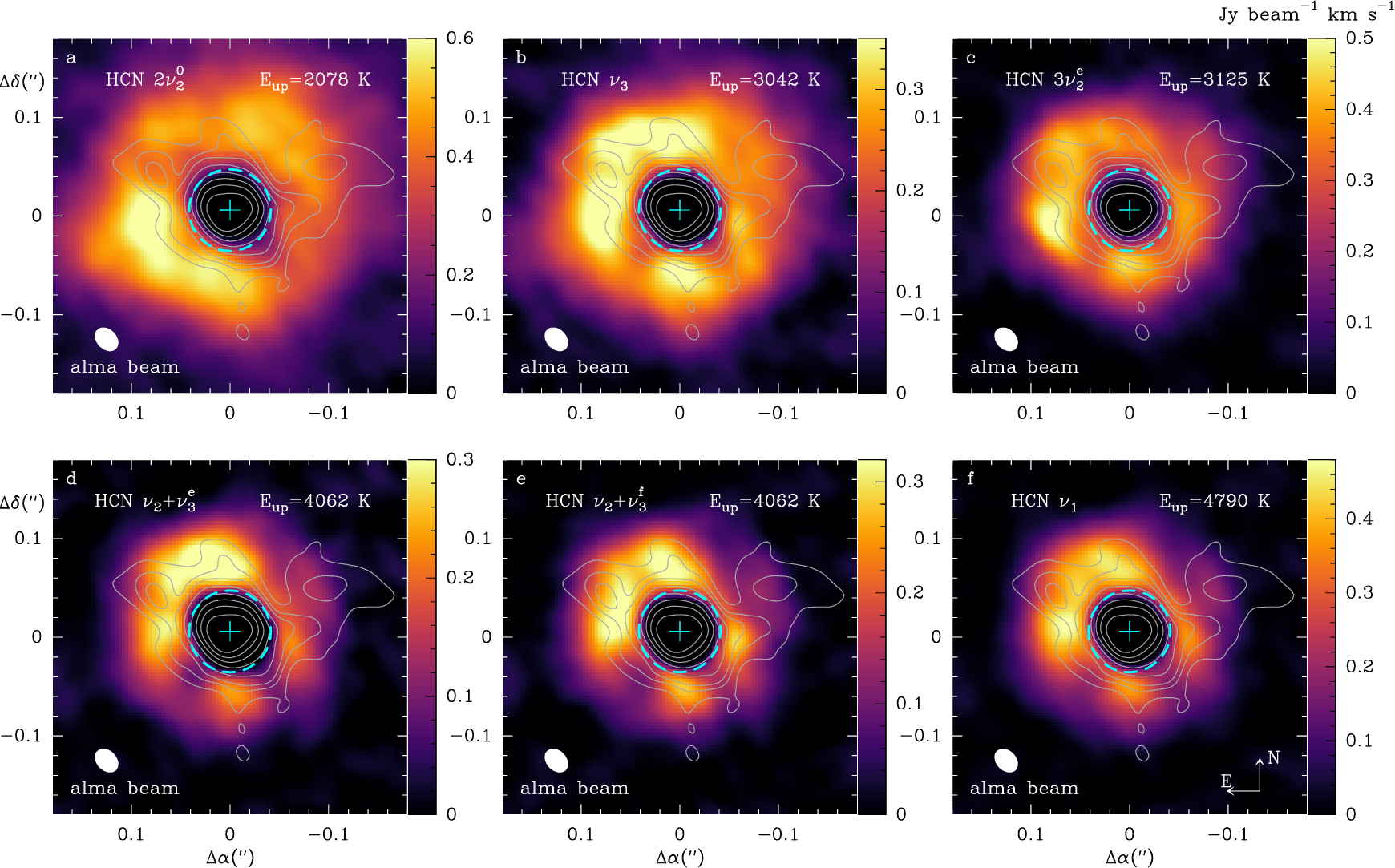}
    \caption{\textbf{Extended data: }Velocity-integrated emission maps of HCN isotopologues in different vibrationally excited states. The colour scale, contours, beam, and orientation are defined as in ED Fig.\,\ref{fig:exdat1}.\label{fig:exdat2}}
\end{figure}
\pagebreak

\begin{figure}
    \centering
    \includegraphics[scale=0.7]{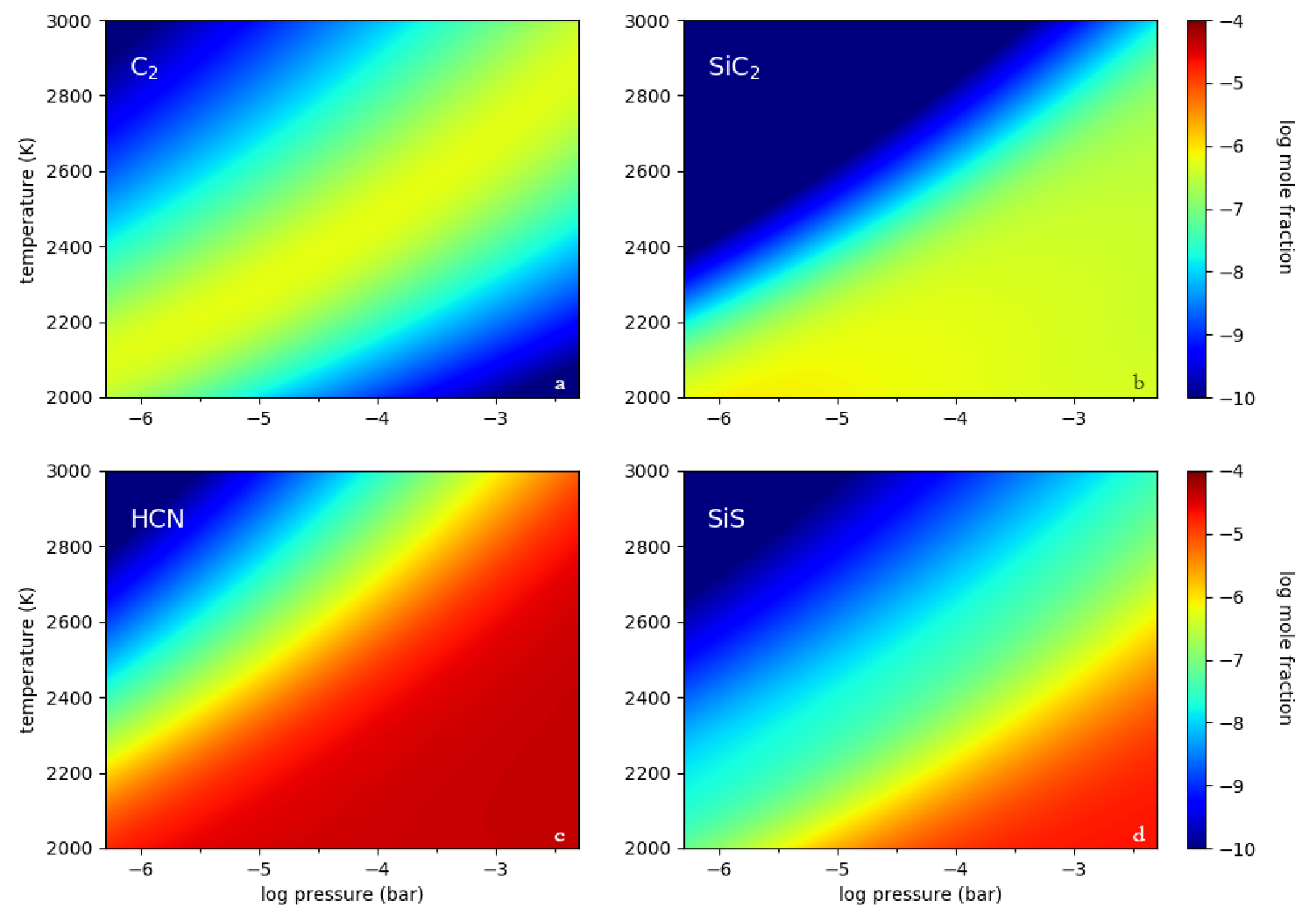}
    \caption{\textbf{Extended data: }Results from the chemical equilibrium models for C$_2$, SiC$_2$, HCN, and SiS. 
    The colour scale map represents the logarithm of the mole fraction of the given molecule as a function of 
    the kinetic temperature and the logarithm of the pressure of the gas.\label{fig:exdat3}}
\end{figure}
\pagebreak

\begin{figure}
    \centering
    \includegraphics[scale=0.73]{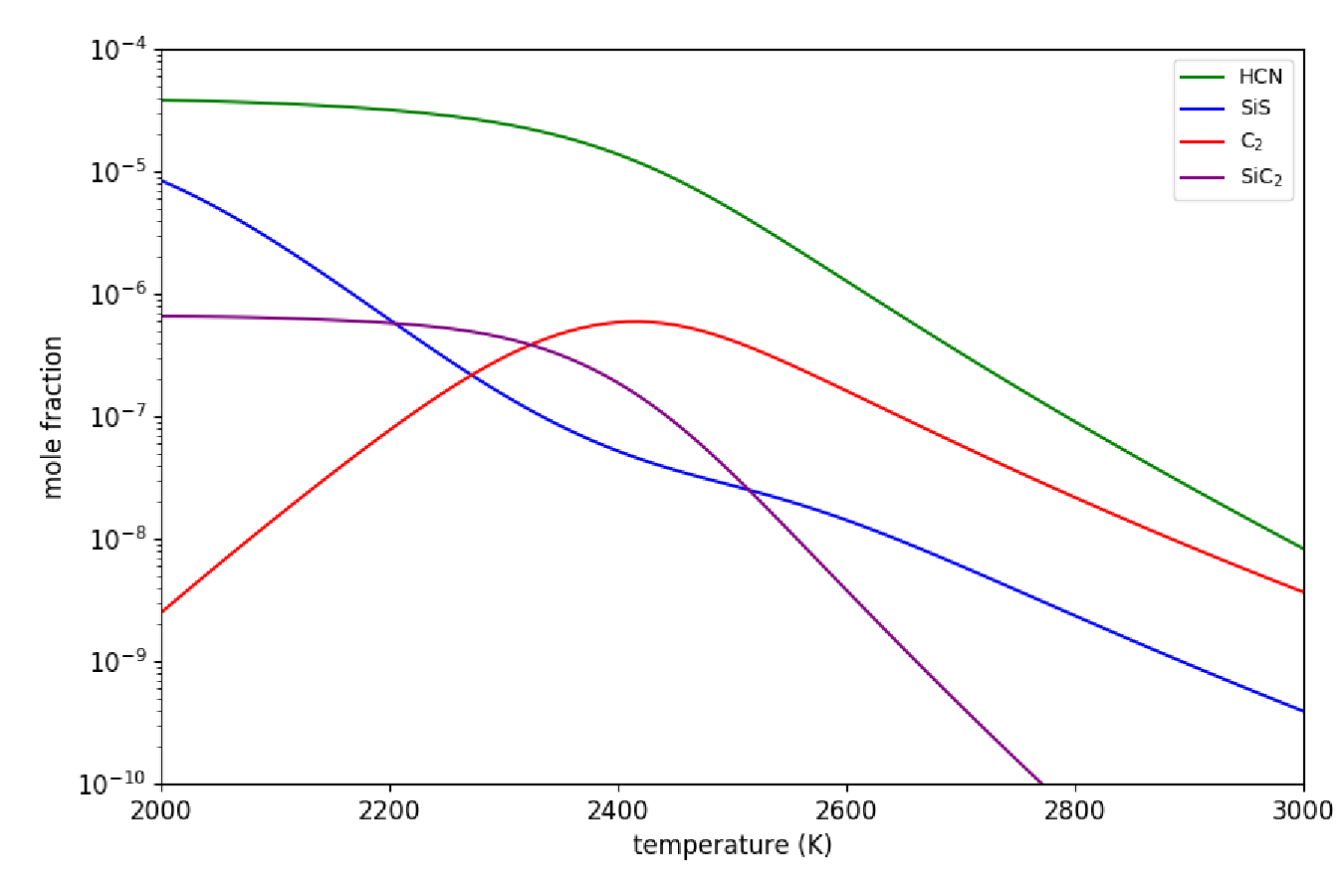}
    \caption{\textbf{Extended data: }Fractional abundance profiles. The abundance profiles are shown as a function of temperature for a fixed pressure of 5\,$\times$\,10$^{-5}$\,bar, which should be close to the gas pressure at the photosphere of \irc\ \citep{2020A&A...637A..59A}.\label{fig:exdat4}}
\end{figure}
\pagebreak

\begin{figure}
    \centering
    \includegraphics[scale=0.63]{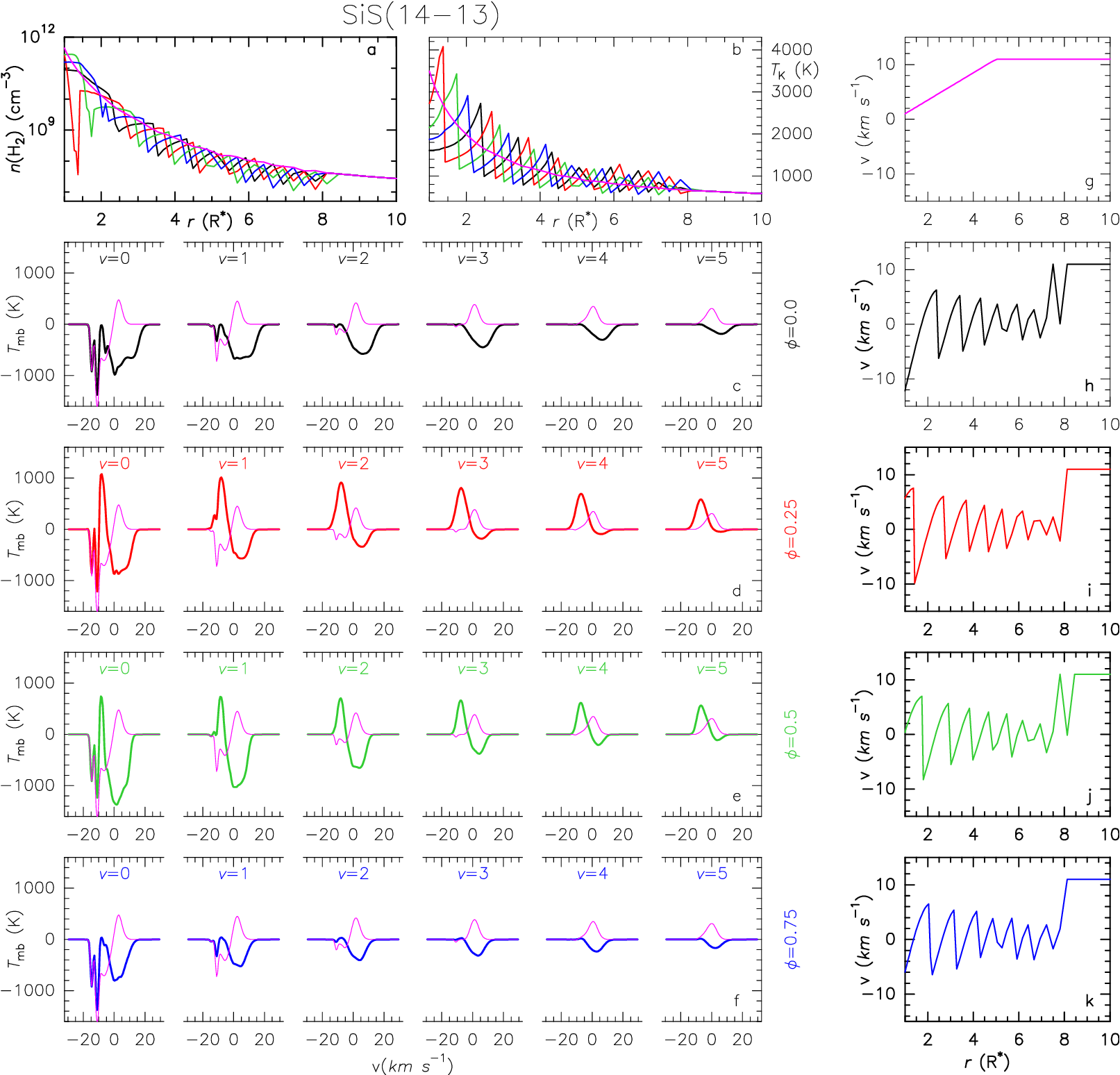}
    \caption{\textbf{Extended data: }Predicted line profiles of the $J$\,=\,14--13 SiS line in different vibrational states as a function of the pulsation phase.
    \textbf{a-b:} density and temperature profiles, where oscillating lines represent the phases $\phi$\,=\,0 (black), $\phi$\,=\,0.25 (red), $\phi$\,=\,0.5 (green), and $\phi$\,=\,0.75 (blue).
    An additional model with no stellar pulsation using density and temperature profiles that falls as $r^{\mathrm{-2}}$ and $r^{\mathrm{-0.7}}$, respectively, is shown as a pink (non-oscillating) solid line.
    \textbf{c-f:} Emergent line profiles where the coloured lines, which follow the colour code mentioned before, represent the predictions for an angular resolution of 0\arcsecp02. 
    \textbf{g-h:} Velocity profile of the gas following also this colour code.\label{fig:exdat5}}
\end{figure}
\pagebreak

\begin{figure}
    \centering
    \includegraphics[scale=0.6]{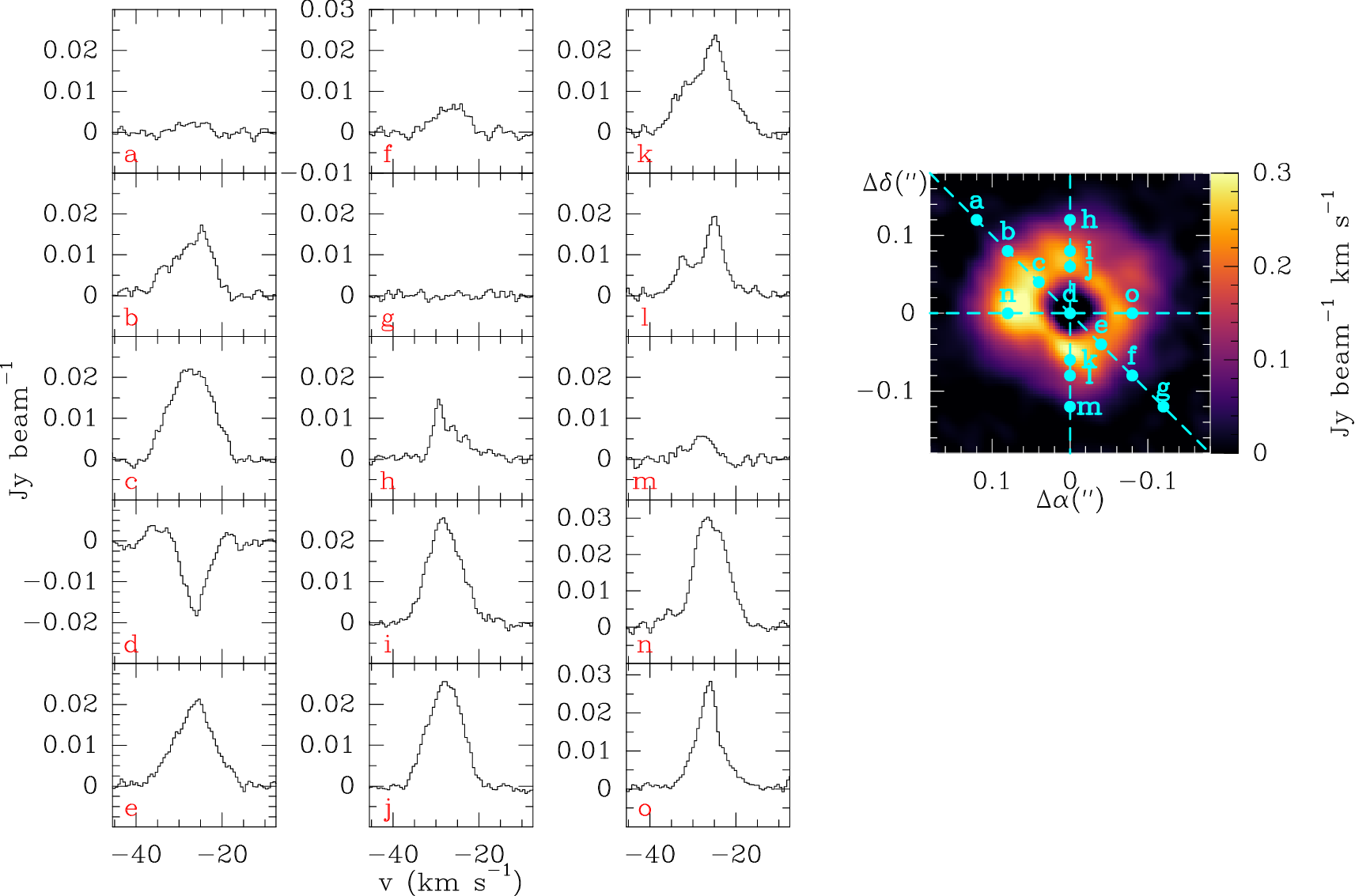}
    \caption{\textbf{Extended data: }Line profiles of the $\nu$\,=\,2 $J$\,=\,14--13 SiS line as observed at different locations in the plane of the sky.
    The lines have been obtained by averaging the spectra in an area of 10\,mas surrounding each selected position. 
    The different numbered offset positions are shown and numbered in the total intensity map to the right. 
    Each corresponding position number is shown in red at the bottom-left corner of the boxes displaying the plots of the spectra.\label{fig:exdat6}}
\end{figure}
\pagebreak

\begin{figure}
    \centering
    \includegraphics[scale=0.6]{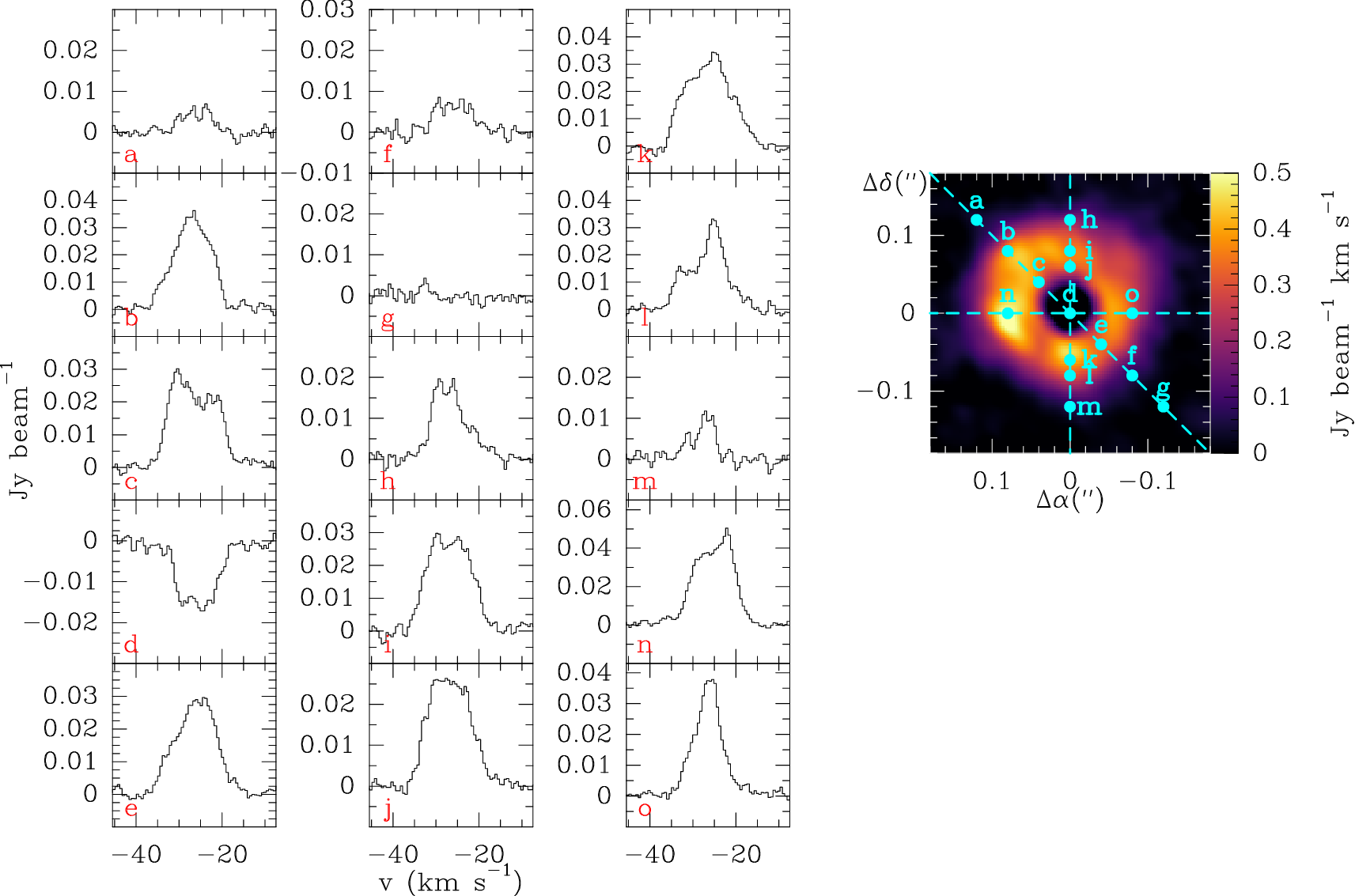}
    \caption{\textbf{Extended data: }Line profiles of the HCN 3$\nu_\mathrm{2}^\mathrm{e}$ $J$\,=\,3--2 line as observed at different locations in the plane of the sky.
Details of this figure are defined as in ED Fig.\,\ref{fig:exdat6}.\label{fig:exdat7}}
\end{figure}
\pagebreak

\begin{figure}
    \centering
	\includegraphics[scale=0.63]{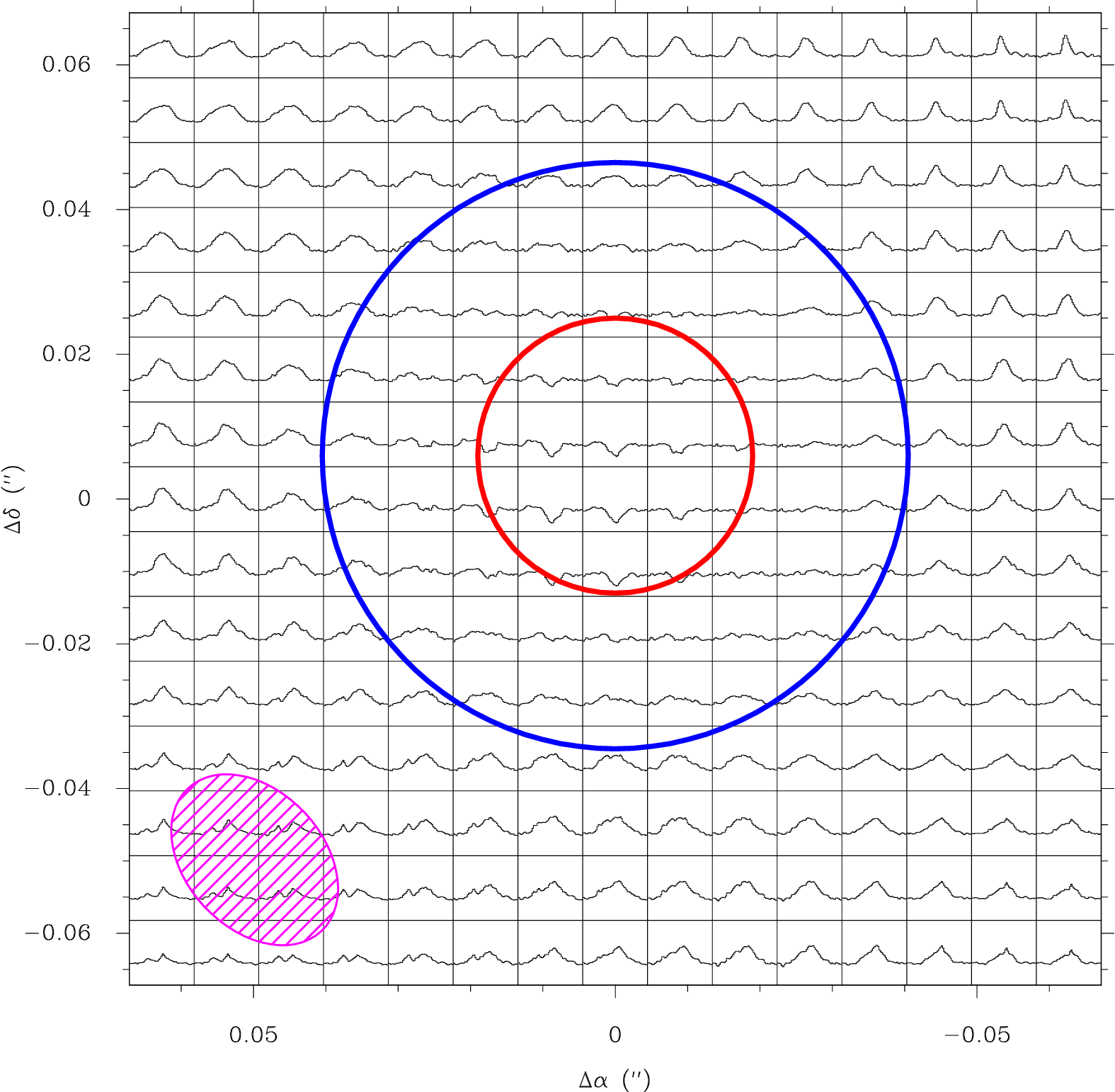}
    \caption{\textbf{Extended data: }Spectra in velocity units at different offsets from the source of the SiS $\nu$\,=\,2 $J$\,=\,14--13  line.
    The velocity range shown for each spectrum is $\pm$\,15\,\kms\ with respect to the systemic velocity of the source, from negative (left) to positive (right).
    The intensity scale, in Jy\,beam$^{-1}$, ranges between -0.03 and 0.06.
    The red circle represents the stellar disk in the IR (radius equal to 19\,mas)\citep{1988ApJ...326..843R}, while the blue circle correspond to the radio photosphere (radius equal to 40.5\,mas)\citep{2012A&A...543A..73M}. \label{fig:exdat8}}
\end{figure}
\pagebreak

\begin{figure}
    \centering
\includegraphics[scale=0.63]{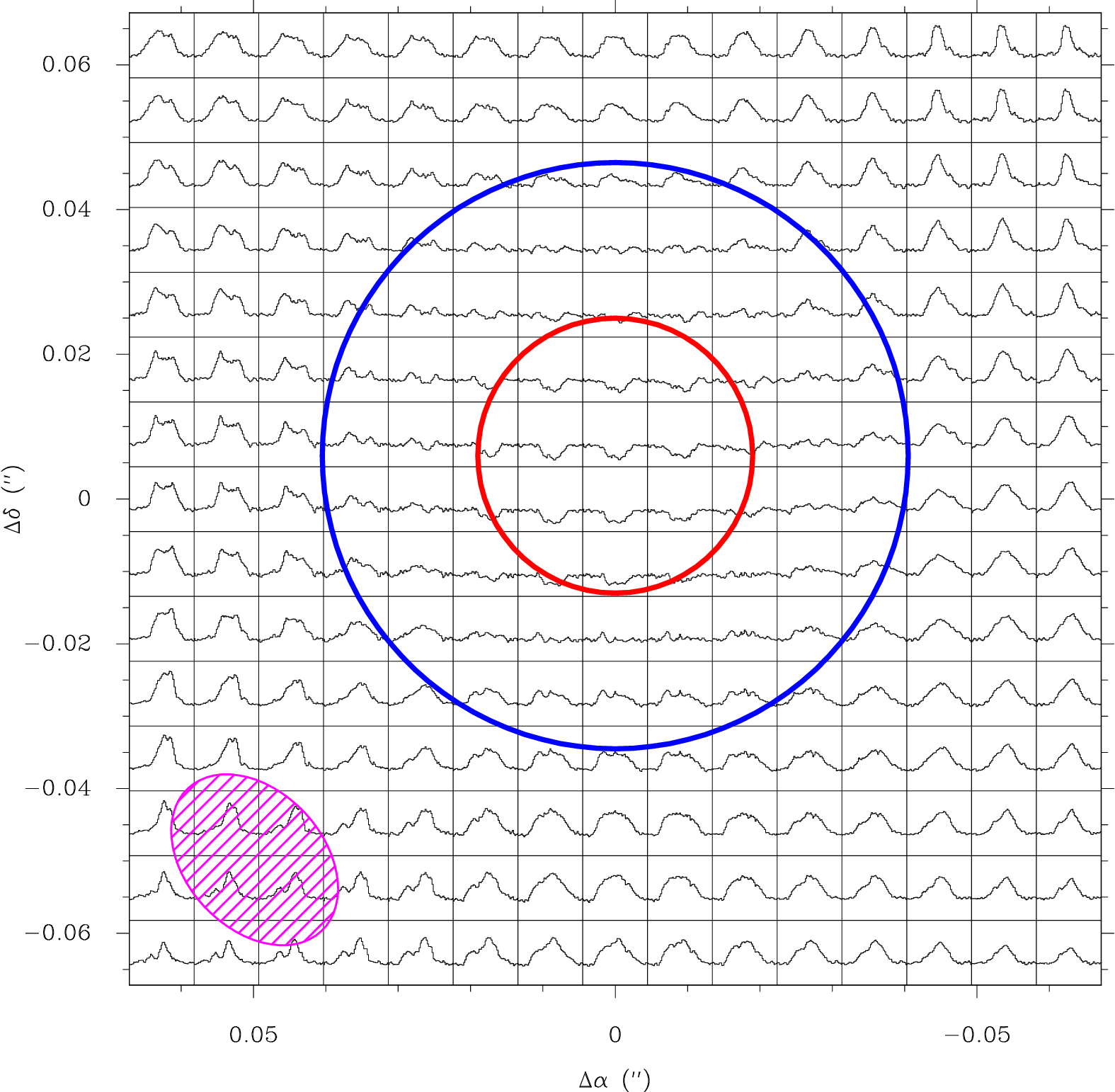}
    \caption{\textbf{Extended data: }Spectra in velocity units at different offsets from the source of the HCN 3$\nu_\mathrm{2}^\mathrm{e}$ line.
Details of this figure are defined as in ED Fig.\,\ref{fig:exdat8}. \label{fig:exdat9}}
\end{figure}
\pagebreak

\begin{figure}
    \centering
    \includegraphics[scale=0.6]{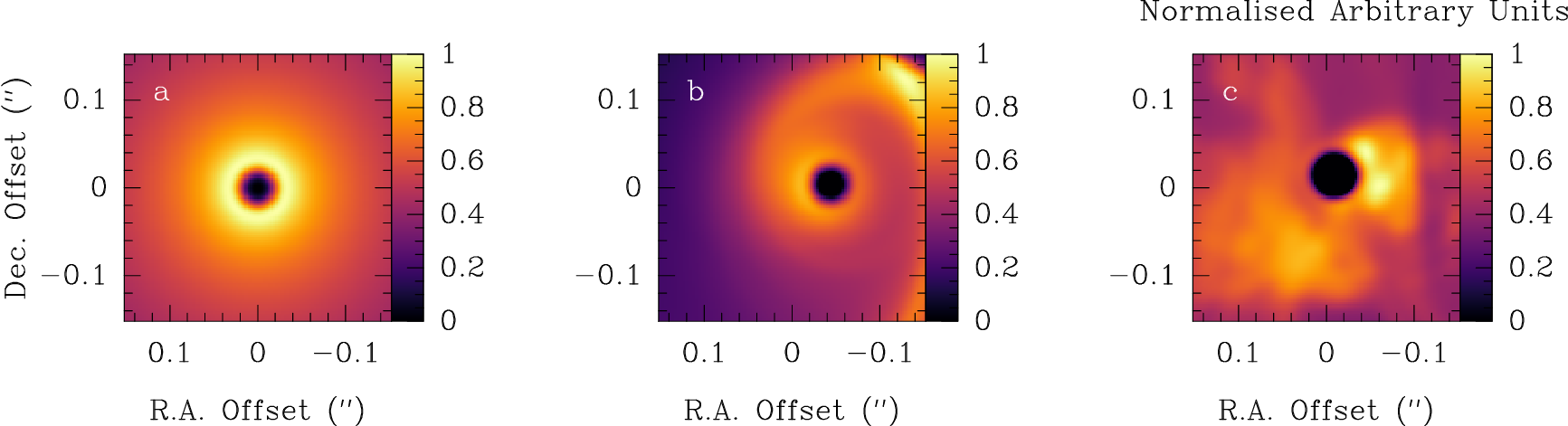}
    \caption{\textbf{Extended data: }Results from the growing envelope model. Three cases are presented: a) control case of an isolated AGB star without formation of blobs, b) AGB steady wind in a binary system with an inclination of 60$^{\circ}$ separated by 25\,AU, and c) as in b but for a closer companion located at 10\,AU and including the randomly generated blobs at the AGB stellar photosphere. The maps are in normalised intensity units to their respective maximum intensity.\label{fig:exdat10}} 
\end{figure}

\end{document}